\documentclass[11pt]{article}

\usepackage{arxiv_pkg}
\usepackage{general_def}

\title{Encoded Gradients Aggregation against Gradient Leakage in Federated Learning}

\author[1,3]{Dun Zeng}
\author[1,3]{Shiyu Liu}
\author[4]{Siqi Liang}
\author[1]{Zonghang Li}
\author[3]{Hui Wang} 
\author[5]{\\Irwin King}
\author[2,3]{Zenglin Xu \thanks{Corresponding author: xuzenglin@hit.edu.cn.}}
\affil[1]{\small{University of Electronic Science and Technology of China}}
\affil[2]{\small{Harbin Institute of Technology, Shenzhen}}
\affil[3]{\small{Peng Cheng Lab}}
\affil[4]{\small{Shenzhen Research Institute, The Chinese University of Hong Kong}}
\affil[5]{\small{The Chinese University of Hong Kong}}

\begin{document}
\date{}
\maketitle

\begin{abstract}
Federated learning enables isolated clients to train a shared model collaboratively by aggregating the locally-computed gradient updates. However, privacy information could be leaked from uploaded gradients and be exposed to malicious attackers or an honest-but-curious server. Although the additive homomorphic encryption technique guarantees the security of this process, it brings unacceptable computation and communication burdens to FL participants. To mitigate this cost of secure aggregation and maintain the learning performance, we propose a new framework called Encoded Gradient Aggregation (\emph{EGA}). In detail, EGA first encodes local gradient updates into an encoded domain with injected noises in each client before the aggregation in the server. Then, the encoded gradients aggregation results can be recovered for the global model update via a decoding function. This scheme could prevent the raw gradients of a single client from exposing on the internet and keep them unknown to the server. EGA could provide optimization and communication benefits under different noise levels and defend against gradient leakage. We further provide a theoretical analysis of the approximation error and its impacts on federated optimization. Moreover, EGA is compatible with the most federated optimization algorithms. We conduct intensive experiments to evaluate EGA in real-world federated settings, and the results have demonstrated its efficacy.
\end{abstract}

\section{Introduction}

\begin{figure}[h]
  \centering  \includegraphics[scale=0.6]{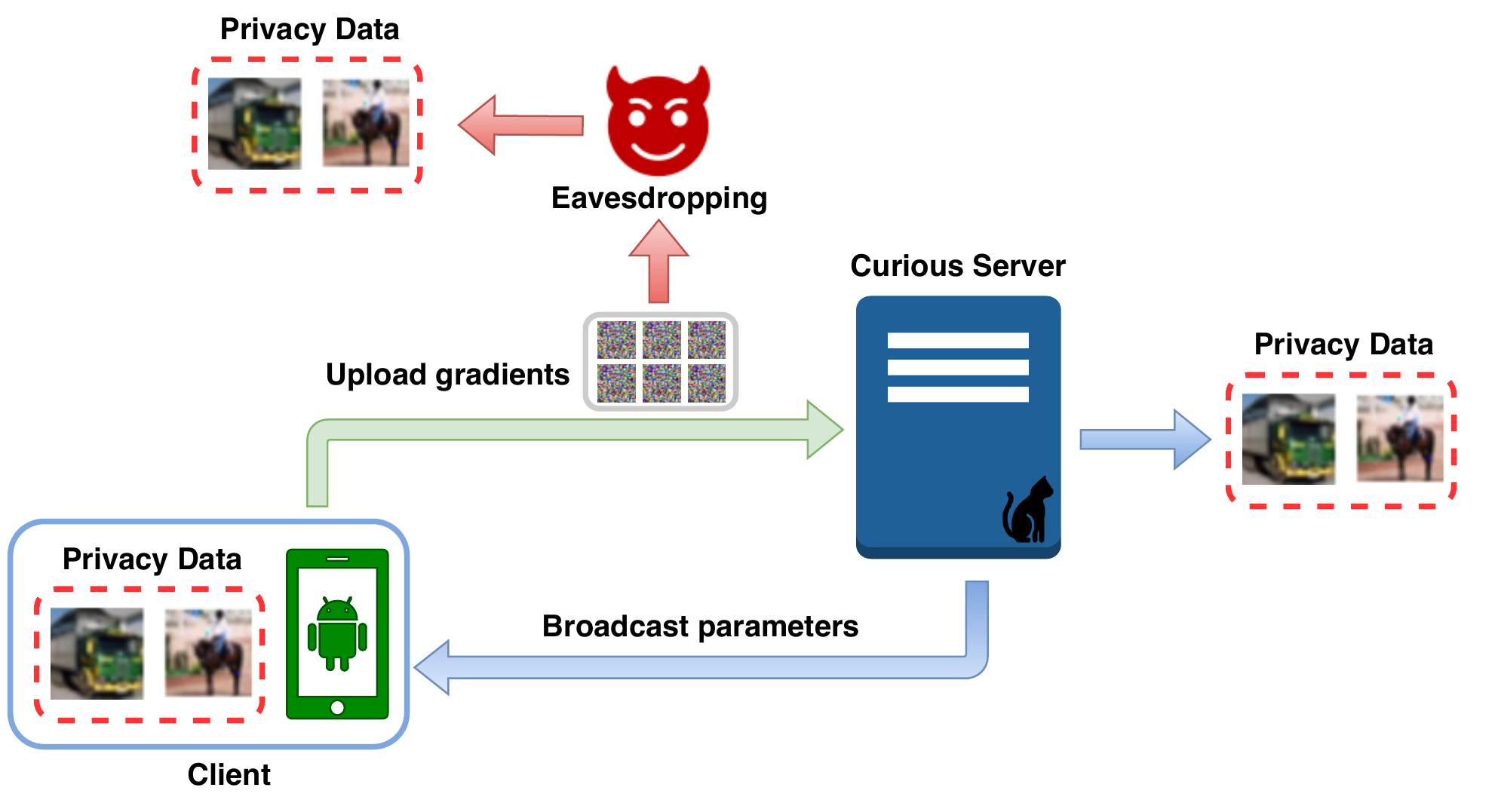}
  \caption{Private data samples are leaked from local updates.}
  \label{fig:mot}
\end{figure}

Federated Learning (FL) is a new distributed learning framework that enables multiple clients to train machine learning models collaboratively without exposing local privacy data~\citep{DBLP:conf/aistats/McMahanMRHA17,kairouz2019advances,DBLP:journals/tist/YangLCT19}. Typically, clients in a federated learning system are required to upload gradients or model parameters to a server that aggregates the gradients or parameters. However, this uploading procedure faces many security vulnerabilities~\citep{DBLP:conf/icml/BhagojiCMC19}. For example, recent studies~\citep{DBLP:journals/corr/abs-2001-02610,DBLP:series/lncs/Zhu020,DBLP:conf/cvpr/YinMVAKM21} suggest that private data samples could be restored from local gradient updates with the gradient inversion technique (see Figure \ref{fig:mot}). Consequently, there is an urgent need of privacy protection techniques for federated learning. 

\begin{figure*}[h]
  \centering
  \includegraphics*[width=\textwidth]{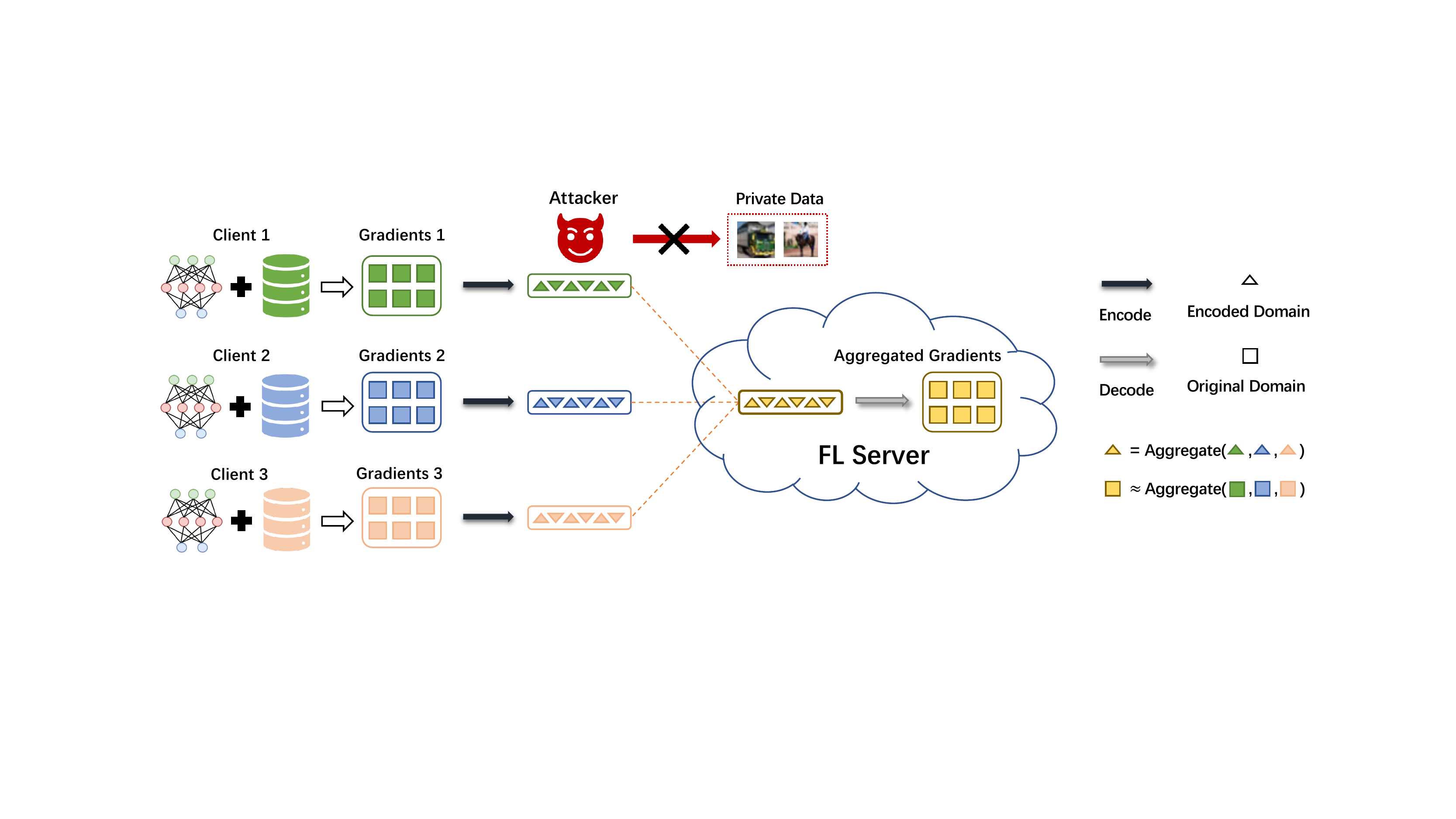}
  \caption{An illustration of EGA in an FL system with 3 clients. The clients upload encoded local gradients to the aggregator; the server decodes the aggregated results and forward the updated global model to clients. 
  }
  \label{fig:orv}
\end{figure*}

A number of techniques have been used to protect the privacy of clients. Multi-Party Computation~\citep{mpcyao, bonawitz2017practical} enables distributed institutions to collaboratively calculate an objective function, which works well in the cross-silo setting (i.e., a few large institutions with sufficient computing and communication resources). Homomorphic Encryption~\citep{gentry2009fully,paillier1999public} allows certain mathematical operations to be performed directly on ciphertexts without prior decryption. In particular, ~\citet{paillier1999public}, an additive homomorphic encryption method, has been implemented in the federated aggregation procedure~\citep{DBLP:conf/usenix/ZhangLX00020, hardy2017private, DBLP:journals/tifs/PhongAHWM18}. Despite the potential advantages of these techniques, they suffer from significant computation and communication costs, impairing the efficiency of FL systems in real-world applications, especially in cross-device scenarios. For instance, when homomorphic encryption is applied in the federated training of the CIFAR dataset~\citep{DBLP:conf/usenix/ZhangLX00020}, compared with directly transferring the plaintext updates, the iteration time of using homomorphic encryption is measured as 2725.7s, which is extended by $135\times$, and the data transfer between clients and aggregator in one iteration is measured as 13.1 GB on average while the number is only 85.89 MB without homomorphic encryption. 
In cross-device federated learning, edge devices cannot afford such tremendous resource consumption. In addition, resource-friendly differential privacy~\citep{wei2020federated, lee2018concentrated, DBLP:conf/ijcai/PhanVLJDWT19} for federated learning could also protect the privacy, however, they may result in significant performance drop. Additionally, Secret Sharing methods~\citep{bonawitz2017practical} provides a way of secure aggregation. However, they are vulnerable to collusion attack.

To mitigate this cost of secure aggregation and maintain the learning performance, we propose a new framework named as Encoded Gradient Aggregation (\emph{EGA}). 
EGA \emph{enables encoded gradients to be aggregated directly for federated learning}. It provides a set of strategies to pretrain an encoder-decoder network for gradient encoding, aggregation, and decoding. The well-pretrained encoder-decoder network can be easily deployed in federated learning procedures as illustrated in Figure~\ref{fig:orv}. In practice, clients hold an encoder network $E_c$ that could encode to-be-upload gradients to an encoded domain with injected noises, which prevents malicious attackers from getting raw gradients (e.g., by monitoring network communication). Then the encoded gradients are sent to the server for direct aggregation, and the server with the decoder network $D_c$ could reconstruct aggregated gradients from the encoded domain. Hence, raw gradients from a single client will not be exposed to the honest-but-curious FL server. In this sense, the designed encoder-decoder process can also be regarded as a numerical approximation framework of homomorphic addition. Importantly, we theoretically prove that the encoding-decoding reconstruction errors brought by EGA are bounded, which ensures that the errors will not impair the model performance while preventing the privacy information leaking. Finally, empirical evaluation has demonstrated that EGA maintains the prediction performance of FL models. Besides, we explore the potential of the encoder-decoder network in communication compression by reducing the length of encoded gradients. 
In summary, the contributions are as follows.
\begin{itemize} 
  \item We propose the EGA framework, which provides an efficient way to securely aggregate gradients in an encoded domain for federated learning. 
  
  \item We theoretically analyze the encoding-decoding error bound, and provide convergence guarantees for EGA. 
  
  \item Intensive evaluation has demonstrated promising performance of EGA in both protecting gradient leakage (as shown in Figure~\ref{tab:dp_atk}) and maintaining the classification accuracy (as shown in Figure \ref{fig:convergence}). Moreover, the communication efficiency study reveals the potential of EGA for saving communication budgets (as shown in Figure~\ref{fig:cmp} ). 
  
\end{itemize}

\section{Related Work}

Recently, a number of federated optimization algorithms have been successfully proposed, including FedAvg~\citep{DBLP:conf/aistats/McMahanMRHA17}, FedProx~\citep{DBLP:conf/mlsys/LiSZSTS20}, q-FedAvg~\citep{DBLP:conf/iclr/LiSBS20},SCAFFOLD~\citep{karimireddy2020scaffold}, FedDyn~\citep{DBLP:conf/iclr/AcarZNMWS21}, FedBABU~\citep{oh2022fedbabu}, etc. For example, 
FedAvg~\citep{DBLP:conf/aistats/McMahanMRHA17} is the standard algorithm aggregating model parameters trained across multiple clients based on the quantities of local data; FedProx~\citep{DBLP:conf/mlsys/LiSZSTS20} adds a proximal term into local loss functions to restrict their updates to be close; q-FedAvg~\citep{DBLP:conf/iclr/LiSBS20} dynamically adjusts the averaging weights for fairness concerns. 
It is important to note that, in these federated optimization algorithms, the aggregation procedure plays a vital role in federated learning.

Despite the successes of federated learning in various applications, it still suffers from significant issues in privacy protection. For example, gradient inversion researches~\citep{DBLP:journals/corr/abs-2001-02610,DBLP:series/lncs/Zhu020,DBLP:conf/cvpr/YinMVAKM21} have demonstrated that malicious attackers can restore private data samples from leaked gradients. This suggests a vital challenge for federated learning: \emph{the protection methods of privacy gradients from malicious attackers and an honest-but-curious server}. Modern cryptography techniques (e.g., Secure Multi-Party Computation~\citep{mpcyao} and Homomorphic Encryption~\citep{gentry2009fully}) could guarantee the security of FL systems. 
However, the enormous resource consumption of deploying these techniques is a major obstacle, especially for resource-limited cross-device scenarios. Therefore, resolving the conflicts between the resource cost and gradients protection remains as an open challenge. To alleviate this problem, we proposed a plug-in framework EGA for weighted gradients aggregation in federated learning, which maintains the performance of standard optimization procedures, while providing economy solutions to gradient protection. 
Our proposed work on encoded gradients aggregation has been built upon recent researches on encoder-decoder networks.
In particular, an autoencoder was designed for extracting gradient correlation in distributed learning workers~\citep{DBLP:journals/corr/abs-2103-08870}. The autoencoder enables workers to transmit similar gradients once for gradient compression. 
However, this method is not applicable to FL systems due to privacy restrictions. \citet{DBLP:conf/dcc/LiH19} presents an autoencoder for gradient encryption, which is unable to process changing magnitude of gradients over the training process. Consequently, it brings unacceptable performance loss to models. Meanwhile, there are no mathematical guarantees presented.
Inspired by these studies, we present several strategies to design encoder-decoder networks for encoded gradients aggregation, such that aggregated results can be accurately reconstructed from the encoded domain. In particular, we provide a theoretical analysis of encoding-decoding errors. 

\section{Encoded Gradients Aggregation}
This section provides the details of EGA. Firstly, we set up the to-be-solved problem. Then we describe the details of the pretraining autoencoder for encoded gradient aggregation. Next, we introduce the procedure for applying the encoder-decoder network to the federated learning algorithm. Finally, we theoretically analyze the mathematical properties of EGA. 

\subsection{Preliminaries}\label{sec:setup}

Without loss of generality, we set up the problem with the vanilla federated learning algorithm(FedAvg), which minimizes finite-sum objectives by averaging local model parameters collected from $m$ clients:
\begin{equation}
\nonumber
  \min_{\boldsymbol{w}} \frac{1}{m}\sum^{m}_{i=1}\lambda_i f_i(\boldsymbol{w}_i; \mathcal{D}_i),
\end{equation}
where $f_i(\boldsymbol{w}; \mathcal{D}_i)$ is the loss of prediction on local dataset $\mathcal{D}_i$. Typically, $\lambda_i$ represents the importance of the $i$-th client. We do not discuss the impact of weights $\lambda$ in this paper, so we set $\lambda = 1$ without preference. 

Each round of the FedAvg procedure runs as follows: first, the server selects a random subset of clients $C^t$ and broadcasts a global model $\boldsymbol{w}^{t-1}$ to them. Next, each client $i \in C^t$ lets $\boldsymbol{w}_i^t = \boldsymbol{w}^{t-1}$ and performs $k$ local stochastic gradient descent epochs with learning rate $\eta$ and local dataset $\mathcal{D}_i$. For the $j$-th epoch, the $i$-th client performs: 
\begin{equation}\nonumber
\boldsymbol{w}^t_i = \boldsymbol{w}^{t}_i - \eta \nabla f_{i,j}(\boldsymbol{w}^{t}_{i}; \mathcal{D}_i).
\end{equation}
Then, the $i$-th client uploads the local gradients $\Delta \boldsymbol{w}_i^t = \frac{\boldsymbol{w}_i^{t} - \boldsymbol{w}^{t-1}}{\eta}$. Finally, the server updates global model parameters by:
\begin{equation}
\label{eq:opt}
  \boldsymbol{w}^t = \boldsymbol{w}^{t-1} + \eta \sum^m_{i=1} \frac{1}{m} \Delta{\boldsymbol w}_i^t.
\end{equation}
To simplify the description, we assume that the gradients $\Delta{\boldsymbol w}$ and model parameters $\boldsymbol{w}$ are denoted by a vector, i.e., $\Delta{\boldsymbol w}, \boldsymbol{w} \in \mathbb{R}^d$. We aim to build an encoding-decoding function to implicitly aggregate the model updates from clients for Equation \ref{eq:opt}. Formally, we let $E_c: \mathbb{R}^d \rightarrow \mathbb{R}^h$ denotes encoding function and $D_c: \mathbb{R}^h \rightarrow \mathbb{R}^d$ denotes decoding function. For an ideal pair of encoding-decoding functions $(E_c^{\star}, D_c^{\star})$, the following equation should hold:
\begin{equation}\nonumber
D_c^{\star}\big(\frac{1}{m}\sum^m_{i=1}E_c^{\star}(\boldsymbol{x}_i)\big) = \frac{1}{m}\sum^m_{i=1}\boldsymbol{x}_i, 
\end{equation}
where $m$ denotes the number of random vector $\boldsymbol{x} \in \mathbb{R}^d$. By deploying the above encoder-decoder into the FedAvg procedure, we can rewrite the optimization problem depicted in Equation \eqref{eq:opt} as:
\begin{equation}\nonumber
  \boldsymbol{w}^t = \boldsymbol{w}^{t-1}+\eta \, D_c^{\star}\big(\frac{1}{m} \sum^m_{i=1} E_c^{\star}(\Delta{\boldsymbol w}_i^{t-1})\big).
\end{equation}
In practice, clients encode local gradients by $E_c^{\star}\big(\nabla f(\boldsymbol{w})\big)$, then send encoded gradients to the FL server. The server collects all encoded gradients from clients and obtains the aggregated gradients from $D_c^{\star}\big( \sum^m_{i=1} \frac{1}{m} E_c^{\star}\big(\nabla f_i(\boldsymbol{w})\big)\big)$ directly without knowing the raw gradients of any single client. 

However, the ideal $(E_c^{\star}, D_c^{\star})$ (e.g., homomorphic addition function) may face a potential attack from honest-but-curious server~\citep{roy2020crypt}, who can easily obtain gradients of certain client $i$ by replacing its encoded results with $\boldsymbol{0}$, like:

\begin{align}
\label{eq:atk}
  \Delta\boldsymbol{w_i} = & D_c^*\big(\frac{1}{m} \sum_{j=1}^m E_c^*(\Delta\boldsymbol{w_j}) )\big) \\
  & - D_c^*\big(\frac{1}{m} (\sum_{j=1, j \neq i}^m E_c^*(\Delta\boldsymbol{w_j}) + E_c^*(\boldsymbol{0}))\big)
\nonumber
\end{align}
To prevent this potential privacy risk, we further improve the target encoding-decoding function with an encoding-decoding error denoted by $\boldsymbol{\nu}$. Hence, we aim to build the following encoding-decoding process:

\begin{equation}
  D_c\big(\frac{1}{m}\sum^m_{i=1}E_c(\boldsymbol{x}_i)\big) = \frac{1}{m}\sum^m_{i=1}\boldsymbol{x}_i + \boldsymbol{\nu}.
\end{equation}

Motivated by the neural network for gradient compression in distributed learning~\citep{DBLP:journals/corr/abs-2103-08870,DBLP:conf/dcc/LiH19}, we present strategies to fit the target encoding-decoding function with autoencoders. To build a practical encoder-decoder for federated learning, we aim to achieve two goals:

\textbf{(1) Error-bounded.}
We concentrate on building an error-bounded encoding-decoding scheme (i.e., the noise $\nu$ is bounded). Consequently, the encoding-decoding noise of $(E_c, D_c)$ should not affect the performance of the FL model. 

\textbf{(2) Adaptability.} As the magnitude of gradients varies with several factors (e.g., model architectures, dataset, and hyper-parameters), the strategies for the encoder-decoder dealing with various magnitudes and distribution of gradients can be tricky.

\subsection{Methodology}\label{sec:framework}
We provide details of building an error-bounded encoder-decoder network $(E_c, D_c)$ for encoded domain aggregation. Firstly, we introduce the quantization technique to limit the input domain for the encoder. Then, we describe how to train an encoder-decoder offline for encoded domain aggregation. Finally, we demonstrate the details of applying EGA to federated learning algorithms. 

\textbf{Quantization}.
QSGD~\citep{alistarh2017qsgd} is a lossy gradients compression algorithm in distributed learning, replacing the element of gradients with a lower number of bits. Unlike quantization for compression, we use quantization to transfer the input domain from $\mathbb{R}^d$ to $\mathbb{Z}^d$. It could fix the magnitude of input vector $\boldsymbol{x}$ before the encoder network encodes it. 

We define the quantization function $ Q(\boldsymbol{x}; s, n) = \boldsymbol{l}$ and 
\begin{equation}
  \boldsymbol{l} \triangleq 
  \begin{cases} \lfloor \frac{\boldsymbol{x}}{n}s \rfloor \quad \text{with probability} \; 1-(\frac{\boldsymbol{x}}{n}s - \lfloor \frac{\boldsymbol{x}}{n}s \rfloor), \\ \lfloor \frac{\boldsymbol{x}}{n}s \rfloor + 1 \quad \text{otherwise},\end{cases}
\nonumber
\end{equation}
where $n$ is the hyperparameter for normalization, $s\geq1$ is the quantization level, $\boldsymbol{x} \in \mathbb{R}^d$ and $\boldsymbol{l} \in \mathbb{Z}^d$. For $k \in [d]$, the $k$-th element $-s \leq \boldsymbol{l}^k \leq s$ is an stochastic integer corresponding with real-value $\boldsymbol{x}^k$. 

We use $\mathbb{Q}^d_s$ denote a quantified integer domain for a better description. For all $\boldsymbol{l} \in \mathbb{Q}^d_s$ and $k \in [d]$, we have $ -s \leq \boldsymbol{l}^k \leq s$. The quantified domain $\mathbb{Q}^d_s$ is a limited integer domain, i.e., $\mathbb{Q}^d_s \subseteq \mathbb{Z}^d$. Therefore, the quantization function maps gradients from a large real number domain $\mathbb{R}^d$ to a quantified domain $\mathbb{Q}^d_s$. Quantization level $s$ and the larger input size $d$ leverage the size of the quantified domain; that is, the higher quantification level $s$ and input size $d$ lead to a larger quantization domain. 

\textbf{Offline-training of Encoder-decoder Networks}.
The quantization function could limit the domain of input gradients to $Q_s^d$. Considering model size $d$ is usually millions, we divide parameters into many block vectors, each with a size of $b$. Hence, we define an encoder network to fit the encoding function $E_c: \mathbb{Q}^b_s \rightarrow \mathbb{R}^{h}$ and a decoder network to fit the decoding function $D_c: \mathbb{R}^{h} \rightarrow \mathbb{R}^b$. We train $(E_c, D_c)$ end-to-end and offline, following the objective function below.
\begin{equation}\label{objective function}
   \min_{E_c, D_c} \; \mathcal{L} \Big(D_c\big(\frac{1}{m} \sum^m_{i=1} E_c(\boldsymbol{l}_i)\big) , \frac{1}{m}\sum^m_{i=1} \boldsymbol{l}_i \Big),
\end{equation}
where $\boldsymbol{l} \in \mathbb{Q}_s^b$ and $\mathcal{L}( \cdot , \cdot)$ is the loss function leveraging the distance between the encoding-decoding outputs and the ground truth values. As the input domain is limited, the encoder-decoder network $(E_c, D_c)$ can be pretrained with a synthetic dataset randomly generated from $\mathbb{Q}_s^b$. Therefore, the performance of the pretrained encoder-decoder would be affected by input size $b$, hidden domain $h$, quantization level $s$, and the number of vectors $m$ (i.e., the number of clients for each FL round). We summarize the above process in Algorithm \ref{alg:pretrain}.

\begin{algorithm}[H]
\caption{Offline training encoder-decoder networks}
\label{alg:pretrain}
\SetAlgoLined
\SetAlgoNoEnd
\IncMargin{1em}
\DontPrintSemicolon
\KwIn{Initialized encoder $E_c$ and decoder $D_c$ with input size $b$, vector number $m$ and quantization level $s$.}  
\KwOut{Well-trained $E_c, D_c$.}

Initial train set $D_{train} = \{l_1, l_2, \dots\}, \text{and} \; l_i \in \mathbb{Q}^b_s$

Initial test set $D_{test} = \{l_1, l_2, \dots\}, \text{and} \; l_i \in \mathbb{Q}^b_s$

\For{T epochs}{
  shuffle train set $D_{train}$;
  
  \For{every $m$ samples (indexed by $i$) from $D_{train}$}{
    loss = $\mathcal{L} \Big(D_c\big(\frac{1}{m} \sum^m_{i=1} E_c(\boldsymbol{l}_i)\big) , \sum^m_{i=1} \boldsymbol{l}_i \Big)$
    
    perform backward and update $E_c, D_c$ with certain batch size;

  }
  
  evaluate the performance of $E_c, D_c$ on test set $D_{test}$, and save the best model (i.e., achieve the lowest test loss) as $\Bar{E_c}, \Bar{D_c}$ .
  
}

return well-trained $\Bar{E_c}, \Bar{D_c}$;
\end{algorithm}

To put it simply, the encoder $E_c$ learns hidden representations (encoding) of integer vector $\boldsymbol{l}$. The decoder $D_c$ measures the aggregated result of $m$ encoded vectors and decodes the corresponding aggregation results. Meanwhile, the encoder-decoder introduces a noise $\boldsymbol{\nu}$ between the decoding results and the ground truth values. If the size of the encoded domain $h$ is small, i.e., $h < b$ in pretrain, we could implement a lossy communication compression scheme. In other words, EGA implicitly enables a tiny real-number vector to encode an integer vector. We study this case in the experiment sections. 



\textbf{Applications in Federated Aggregation}. 
Given a well-pretrained encoder-decoder network $D_c, E_c$, quantization function $Q$ and its corresponding parameters $n,s$, we could implement encoded gradient aggregation in federated learning. Formally, we have:
\begin{equation}\label{eq:EGA}
  \frac{n}{s} D_c\Big(\frac{1}{m} \sum^m_{i=1} E_c(Q(\Delta{\boldsymbol w}_i; s, n))\Big) = \sum^m_{i=1} \Delta{\boldsymbol w}_i + \frac{n}{s} \boldsymbol{\nu}.
\end{equation}
Then, we apply the EGA to a general federated learning framework, where the details in Algorithm~\ref{alg:fedcda}. EGA is compatible with the weighted average optimization pattern in FL. The local training procedure is unfixed in Line 18. Besides, the weights for aggregation could be adjusted in Line 20. Hence, we note that EGA can be applied to most weighted average-based algorithms such as FedProx, q-FedAvg, etc.


In practice, the $i$-th client updates its local model on the local dataset $\mathcal{D}_i$, and then quantifies and encodes local gradients $\Delta{\boldsymbol w}_i$. Finally, each client sends its encoded gradients $\boldsymbol{g}_i$ to the server. The server collects all encoded gradients from clients and aggregates them directly in an encoded domain. Then, the server decodes gradients $\boldsymbol{g}_{rec}$ and updates the global model accordingly.

\begin{algorithm}[H]
\caption{Federated learning with EGA}
\label{alg:fedcda}
\SetAlgoVlined
\IncMargin{1em}
\KwIn{Initial model $\boldsymbol{w}^0$, well-pretrained autocoders ($E_c, D_c$), communication round $T$, client set $C$, client dataset $\mathcal{D}$, quantization level $s$ and quantization function $Q(\cdot ; s, n)$}
\KwOut{Federated model $w^T$.}
\SetKwFunction{FMain}{ServerProcedure}
\SetKwFunction{CMain}{ClientUpdate}

\SetKwProg{Pc}{}{:}{}
\SetKwProg{Fc}{}{($w^t, s, n$):}{}
\Pc{\FMain}{
  \For{$t \in [T]$}{
    choose a subset of client $C^t \subseteq C, m = |C^t|$ \\
    \For{$i \in C^t$ \text{in parallel}}{
      $\boldsymbol{g}_i \leftarrow \text{ClientUpdate}(\boldsymbol{w}^t, s, n)$
    }
    
    $\boldsymbol{g}_{rec} \leftarrow \frac{n}{s} D_c(\sum_i^m \boldsymbol{g}_i) $
    
    $\boldsymbol{w}^{t+1} \leftarrow \textrm{GlobalUpdate}(\boldsymbol{w}^t, \boldsymbol{g}_{rec})$
  }
}

\Fc{\CMain}{
  $\boldsymbol{w} \leftarrow \boldsymbol{w}^t $ 
  
  $\boldsymbol{w} \leftarrow \textrm{LocalTraining}(\boldsymbol{w}; \mathcal{D}_i)$
  
  $\Delta{\boldsymbol{w}} \leftarrow \boldsymbol{w} - \boldsymbol{w}^t$
  
  $\boldsymbol{g} \leftarrow E_c\big(Q(\Delta{\boldsymbol{w}} \cdot$ $\lambda_i$ $; s, n)\big)$
  
  return $\boldsymbol{g}$
}

\end{algorithm}

\subsection{Mathematical Analysis}\label{sec:math}



\textbf{Bound of Error}. Here we discuss the overall error $\boldsymbol{\nu}$. The input of the encoder-decoder network is quantified. It adds an encoding-decoding error to the results in the quantization domain $Q_s^d$. Hence, the overall error includes the quantization error and encoding-decoding error. And the overall error $\boldsymbol{\nu}$ will be further adjusted, i.e., $\frac{n}{s} \boldsymbol{\nu}$ term in Equation \eqref{eq:EGA}. We study its impacts and present the variance bound of EGA in Theorem \ref{bound:cda} and provide the proof in Appendix.


\begin{theorem}\label{bound:cda}
For all $x \in \mathbb{R}^d$, the variance of $k$-th encoded aggregation $\|\frac{n}{s} D_c(\frac{1}{m} \sum^m_{i=1} E_c(Q(\boldsymbol{x}_i^k; s, n))) - \frac{1}{m}\sum_{i=1}^m \boldsymbol{x}_i^k\|_2^2$ is bounded by $\frac{n}{m}\min(\frac{d}{s^2}, \frac{\sqrt{d}}{s})+ \frac{n^2}{s^2}\sigma^2$, where the $\sigma$ denotes the expectation of encoding-decoding error brought by $(E_c, D_c)$.
\end{theorem}

Theorem \ref{bound:cda} demonstrates that the variance bound is effected by the normalization parameter $n$, quantization level $s$ and the number of clients $m$. Furthermore, we can increase $m, s$ and reduce $n$ for the lower magnitude of the noise.

\textbf{Convergence Analysis.} 
Without loss of generality, we analyze the convergence performance of EGA when it applies to FedAvg. Given access to stochastic gradients, and a starting point $\boldsymbol{w}^{0}$, build iterates $\boldsymbol{w}^{t}$ given by Algorithm \ref{alg:fedcda}. In this setting, we have:
\begin{theorem}\label{them:conv}
Let $\mathcal{W} \subseteq \mathbb{R}^{d}$ be convex, and let $f: \mathcal{W} \rightarrow \mathbb{R}$ be unknown, $\mu$-strongly convex, and $L$-smooth. Let the variance of the stochastic gradient in each clients is upper bounded $\mathbb{E}\Big[\big\|\nabla f_{i}\big(\boldsymbol{w}\big)\big\|_{2}^{2}\Big] \leq G^{2}$. Let $\boldsymbol{w}^{0} \in \mathcal{W}$ be given, and let $R^{2}=\sup _{w \in\mathcal{W}}\big\|\boldsymbol{w}-\boldsymbol{w}^{0}\big\|^{2}$. Let $T>0$ be fixed. Considering $\eta_t=\eta\leq \min \big\{1, \frac{1}{\mu k}\big\}$ and $k=1$, we have
$$
\begin{aligned}
\mathbb{E}[f(\boldsymbol{w}^{T})]-&\min _{\boldsymbol{w} \in \mathcal{W}} f(\boldsymbol{w}) \leq \frac{L}{2}(1-\mu \eta)^{T}R^2 \\
&+\frac{\eta L}{2\mu}\Big(G^{2}+ \frac{n^2}{s^2}d\sigma^2\Big)\big(1-(1-\mu \eta)^{T}\big).
\end{aligned}
$$
\end{theorem}

Theorem \ref{them:conv} is convergence analysis results when $k=1$ (i.e., each client performs once gradient descent locally). For more general conclusions, please see our proof in Appendix. Theorem \ref{them:conv} demonstrates the factors affecting the optimization performance of FedAvg with EGA. It suggests choosing a larger quantization level $s$ and a well-pretrained encoder-decoder with lower $\sigma$. The normalization parameter $n$ is supposed to decay over the training process for better convergence results.

\section{Experimental Evaluation}\label{sec:exp}

In this section, we first study the factors affecting offline pre-trained encoder-decoder networks on synthetic datasets. Then, we evaluate the evaluate performance of EGA on real-world datasets. Finally, we discuss the potential improvements of EGA in real-world applications. We implement our experiments via a open-source federated learning framework~\citep{zeng2021fedlab}.

\subsection{Pretraining of Encoder-decoder Network}

The encoder-decoder networks are symmetrical in architecture in this paper, and each of them is a multilayer perceptron with residual blocks and ReLu activation. The size of them is less than 30 MB. Without loss of generality, we empirically study the properties of the proposed strategies for pretraining such encoder-decoder networks. We train them with $10,000 * m$ randomly generated vectors and test their encoding-decoding performance with another $5,000 * m$ randomly generated vectors. We use the MSE loss for the pretrain target in Equation \eqref{objective function} and the Adam optimizer to utilize the training process. We report the standard variance of an encoding-decoding error on the test set and observe its trend with hyper-parameters change.

\begin{figure}[h]
  \centering
  \subfigure[$b=256, h=128$]{\includegraphics[width=0.4\textwidth]{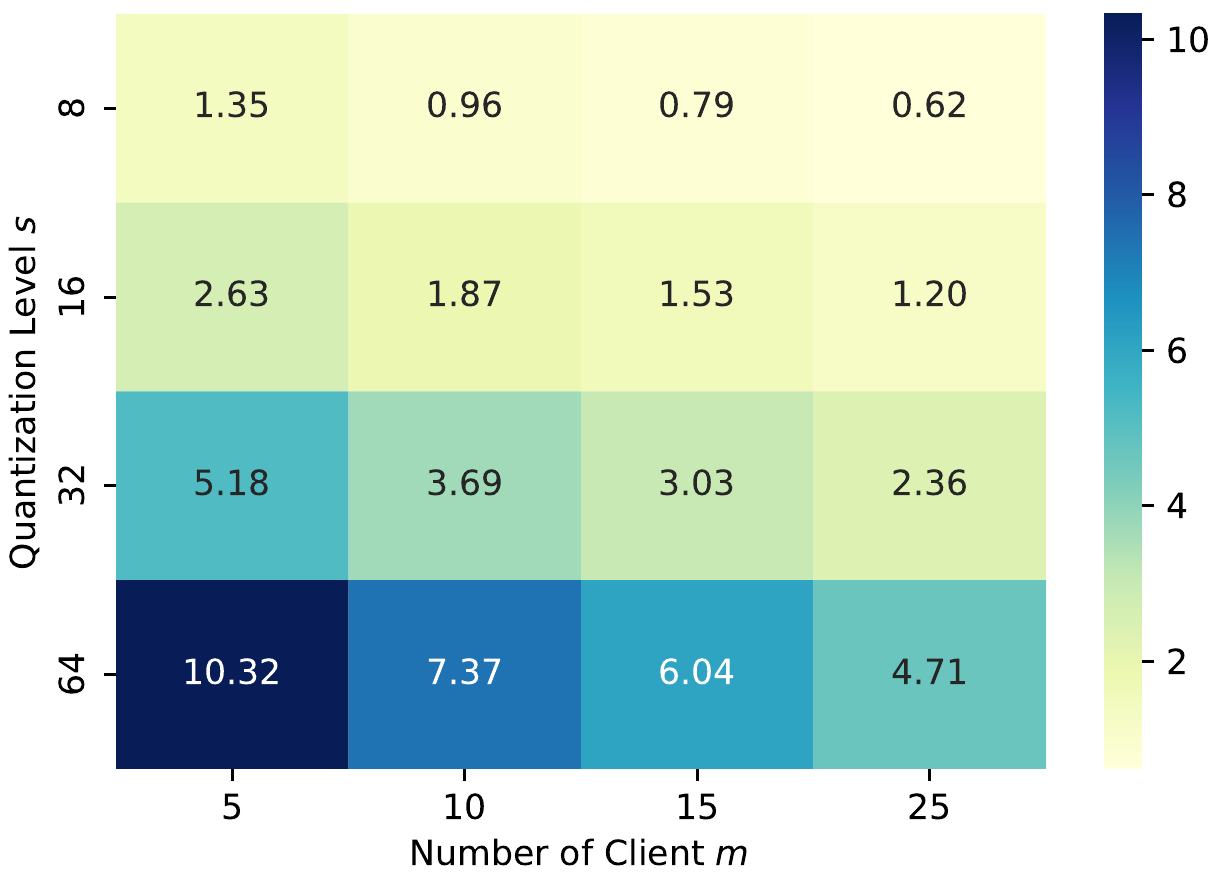}}
  \vspace{4pt}
  \subfigure[$m=10, s=64$]{\includegraphics[width=0.4\textwidth]{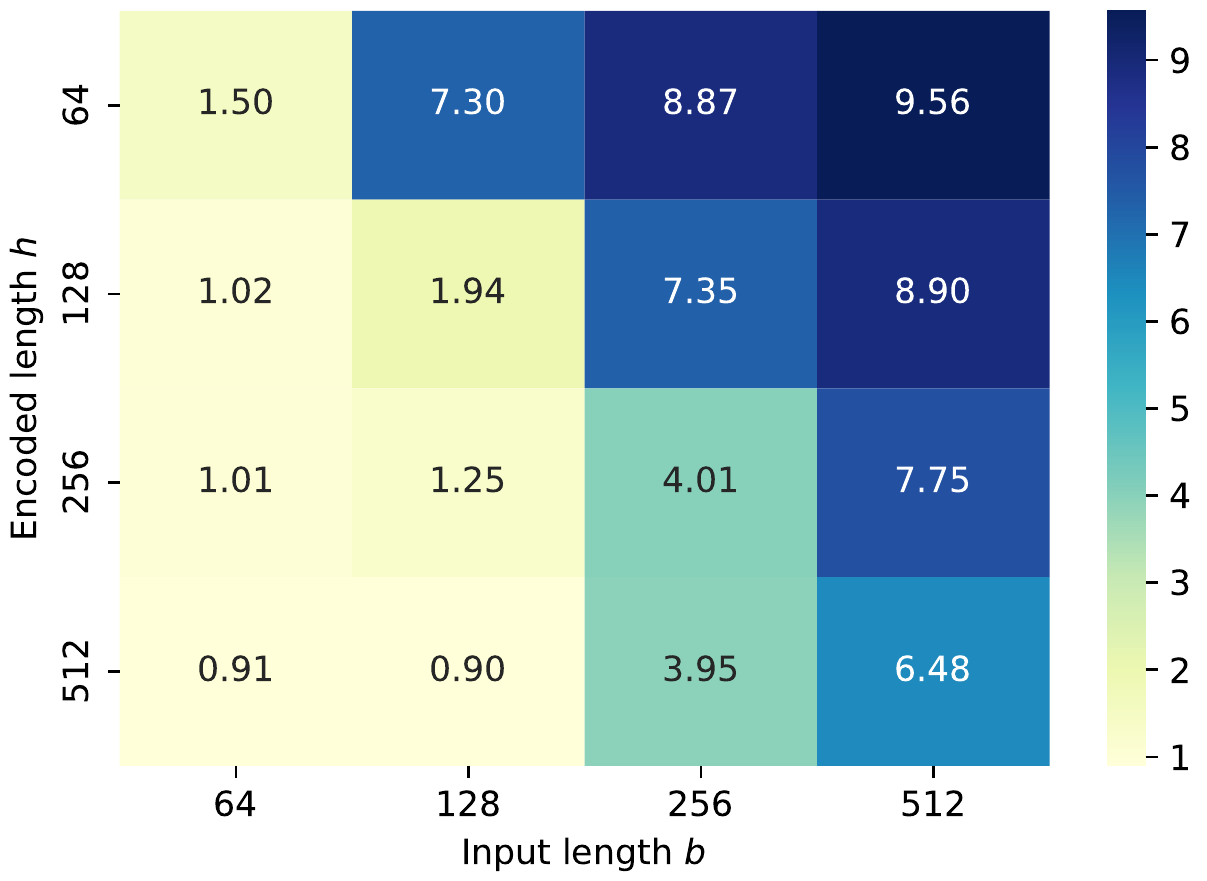}}
	\caption{Fix $b=256, h=128$, then we train encoder-decoder with $m=\{5,10,20,25\}$ and $s=\{8, 16, 32, 64\}$. Fix $m=10, s=64$ to observe the test error under different $b,h \in \{64, 128, 256, 512\}$.}
	\label{fig:pretrain}
\end{figure}

\begin{figure*}[h]
  \centering
  \includegraphics[width=0.9\textwidth]{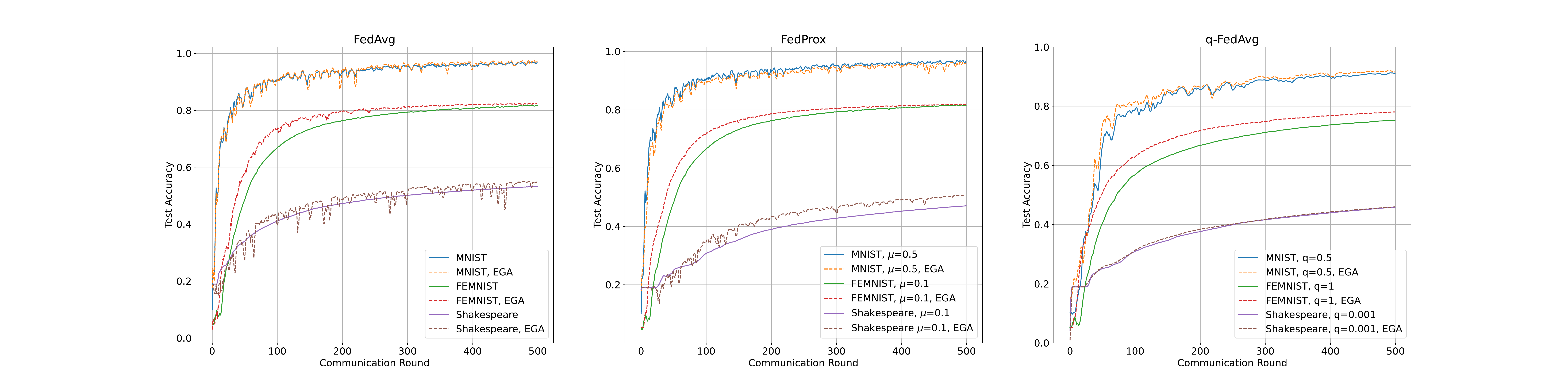}
	\caption{ EGA with different federated optimization algorithms.}
	\label{fig:convergence}
\end{figure*}

We pretrain encoder-decoder networks with 100 epochs following Algorithm \ref{alg:pretrain}. As shown in Figure \ref{fig:pretrain}, the results support the Theorem \ref{bound:cda}, which concludes that the hyper-parameters $m,b,s$ affect the upper bound of the encoding-decoding error. The higher $m,s$ and lower $b$ for the pre-train achieve a lower encoding-decoding error. The experiment results are consistent with previous theoretical analysis. These results suggest that the encoder-decoder network can achieve various noise levels in the quantization domain by properly choosing pretrain hyper-parameters.

\subsection{Evaluation Experiments on EGA}\label{sec:performance}

\textbf{Experimental Settings}.
To show the adaptability of EGA, we evaluate its practical performance on several federated learning tasks: image classification on MNIST~\citep{deng2012mnist} and FEMNIST datasets, next-word-prediction on Shakespeare~\citep{shakespeare2014complete,DBLP:journals/corr/abs-1812-01097} dataset. We evaluate EGA on different neural networks. Specifically, we use a linear classifier for the MNIST task, a CNN for the FEMNIST task, and an RNN for the Shakespeare task. For the local updates procedure on the client, we choose SGD optimizer for the clients in MNIST/FEMNIST tasks and ADAM optimizer for Shakespeare clients.
For the \textbf{federated data partition}, we sort the MNIST dataset samples by labels and split them into 200 shards. Then, we assign two shards to each different client~\citep{DBLP:conf/aistats/McMahanMRHA17}. For the Shakespeare and FEMNIST dataset, we follow the partition of the FL benchmarks LEAF~\citep{DBLP:journals/corr/abs-1812-01097}, which partitions (Non-IID) Shakespeare into 660 clients and FEMNIST into 3597 clients. The number of data samples is unbalanced among clients. The data partition information are summarised in Table \ref{tab:data}.

\begin{table}[h]
\centering
\caption{Statistics of real federated datasets}
\label{tab:data}
\begin{tabular}{lrrrr}
\toprule
\multicolumn{1}{c}{\multirow{2}{*}{\textbf{Dataset}}} & \multirow{2}{*}{\textbf{Devices}} & \multirow{2}{*}{\textbf{Samples}} & \multicolumn{2}{c}{\textbf{Samples/device}} \\
\multicolumn{1}{c}{}             &             &             & mean       & std       \\ \hline
MNIST                    & 100           & 60,000         & 600      & 0         \\
FEMNIST                   & 3,597          & 734,590         & 204      & 70         \\
Shakespeare                 & 660           & 3,678,451        & 5,573     & 6,460         \\ \bottomrule
\end{tabular}
\end{table}


\textbf{Impacts on Federated Optimization}. We pretrain three encoder-decoder networks for the number of clients $m = 10, 40, 100$ with the same quantization level $s=64$, input vector length $b=256$, encoded vector length $h=256$. We evaluate EGA with three common federated optimization algorithms, including FedAvg, FedProx, and qFedAvg. FedAvg is a basic federated optimization algorithm. FedProx improves FedAvg by modifying the local training process on clients to address the heterogeneous issue. Hence, FedProx is included at line 13 in Algorithm \ref{alg:fedcda}. q-FedAvg achieves fairness by adjusting the aggregation weights of each client. EGA implements q-FedAvg via changing $\lambda_i$ at line 20 in Algorithm \ref{alg:fedcda}. According to our analysis on Theorem \ref{them:conv}, we dynamically set $n$ as the largest absolute value of gradients for each round. Typically, $n$ decays over the process of training. 

The convergence performance of EGA is shown in Figure \ref{fig:convergence}. The convergence curve indicates that federated optimization algorithms with EGA follow the basic trend with those without EGA. Furthermore, the encoding-decoding error(noise) could improve the model performance in this case. This phenomenon is studied by \citep{DBLP:journals/corr/NeelakantanVLSK15, simsekli2019tail, wu2020noisy}, which concludes that adding noise to the gradient during training helps the model generalize better. Hence, we conclude that the noise of vanilla EGA will not damage the model performance of federated learning.

\begin{figure}[h]
  \centering
  \includegraphics[scale=0.45]{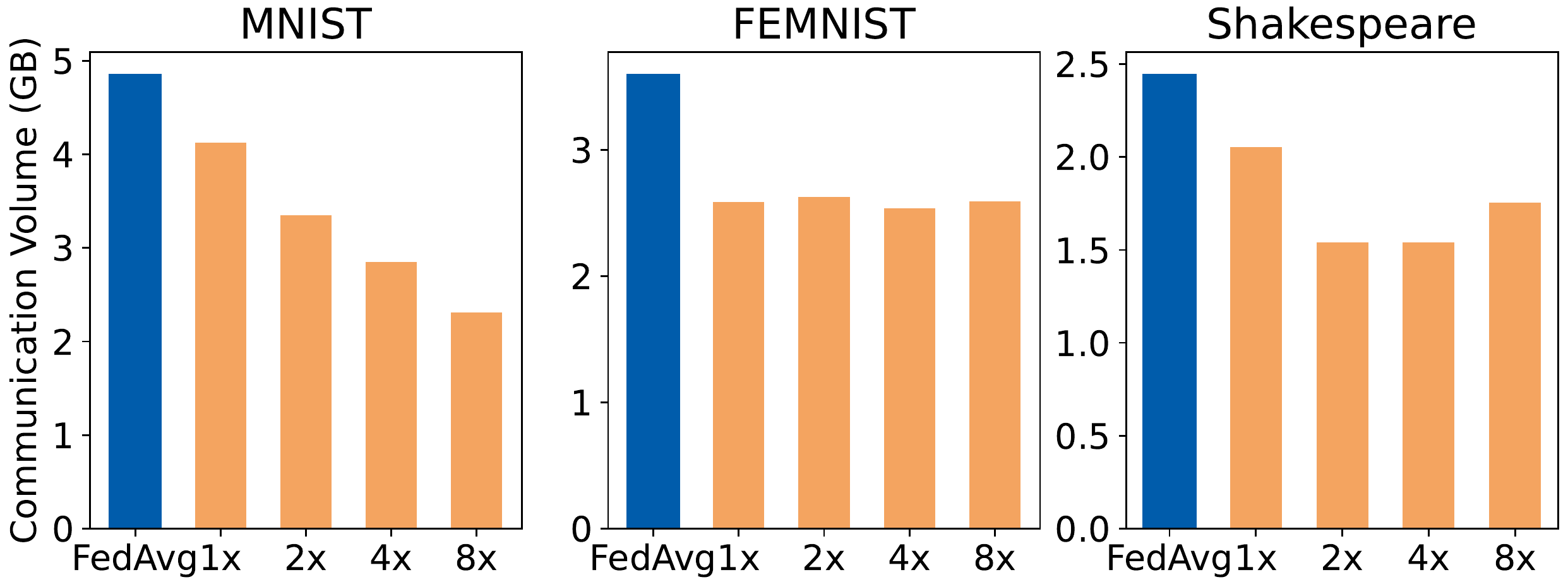}
  \caption{Total client-to-server communication volume.}
  \label{fig:cmp}
\end{figure}

\textbf{Communication Efficiency}. An extra benefit provided by EGA is communication compression. A client could compress their local encoded updates for each round. We study the communication compression potential of EGA. We pretrain encoder-decoder networks with a fixed input length $b=\{512\}$ and encoded length $h=\{256, 128, 64\}$, resulting in different communication efficiency levels marked by $2\times, 4\times, 8\times$. We deploy them on the FedAvg task, respectively. We study the number of communication rounds EGA would cost to achieve the target accuracy. Firstly, we run 100 rounds on FedAvg to search for the best accuracy and set it as the target accuracy for EGA. Then, we run EGA till it first achieves the target accuracy. The results are summarized in Table \ref{tb:ce}, and its corresponding communication cost is shown in Figure \ref{fig:cmp}. Clients could obtain this communication efficient benefits when uploading its encoded local updates. However, the pretrained encoder with a high compress level introduces looser $\sigma$ to aggregated results on the server side. This error causes federated learning to cost more communication rounds to achieve a certain target according to Table \ref{tb:ce} and Theorem \ref{them:conv}. Figure \ref{fig:cmp} suggests a trade-off for an FL system with EGA. A server could spend more downlink communication costs for less uplink bandwidth and volume cost on the client side, where the cost of uplink bandwidth is more expensive.


\begin{table}[h]
\centering
\caption{Communication round to achieve target accuracy.}
\label{tb:ce}
\begin{tabular}{lrrr}
\toprule
Setting & MNIST & FEMNIST & Shakespeare \\ \hline
EGA 1$\times$   & 85     & 72      & 84         \\
EGA 2$\times$    & 138     & 146      & 126        \\
EGA 4$\times$    & 235     & 282      & 252        \\
EGA 8$\times$    & 381     & 577      & 575        \\ \bottomrule
\end{tabular}
\end{table}

\textbf{Prevent Gradient Leakage}. We evaluate whether EGA could defend against gradient inversion attacks in this part. The experiment setting of this part follows DLG~\citep{DBLP:journals/corr/abs-2001-02610}, in which the base model is LeNet-5~\citep{lecun2015lenet} with the Sigmoid activation function, and the data samples are from CIFAR10\citep{krizhevsky2009learning}.
We implement an FL group with ten clients and EGA (the encoder-decoder network is the same as the MNIST experiment). Ten clients perform one step of gradient descent with a single data point, respectively, which is the most vulnerable setting. We obtain the leaked gradient from the view of an honest-but-curious server (described in Equation \ref{eq:atk}). Then, we perform the DLG procedure on the obtained gradients, respectively. The visualization of the gradient inversion procedure is shown in Table \ref{tab:dp_atk}. DLG suggests that a sufficiently large magnitude of noise prevents deep leakage. Therefore, the results indicate that the noise of EGA can prevent deep leakage.

\subsection{Discussion \& Future work }
\textbf{Details of Encoder-decoder Networks}. In practice, the encoder-decoder network should be trained and presented by a trusted party, who delivers the decoder to the FL server and the encoder to clients. Hence, this requires once minority downlink communication cost for clients. In this paper, the specific size of the encoder is relatively tiny. For instance, the encoder size (used in MNIST) is $27$ MB with $6,824,704$ parameters. The additional computation on clients is the forward calculation of encoding whole model gradients. 

\begin{table}[h]
\caption{The visualization of gradient inversion. The left column indicates the iteration epoch of DLG.}
\begin{center}
\renewcommand{\arraystretch}{2}
\begin{tabular}{c|cc|cc|}
\label{tab:dp_atk}
Iter & Raw     & Ours     & Raw     & Ours     \\ \midrule
0     
& \raisebox{-.5\height}{\includegraphics[scale=1.3]{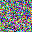}}       
& \raisebox{-.5\height}{\includegraphics[scale=1.3]{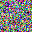}}       
& \raisebox{-.5\height}{\includegraphics[scale=1.3]{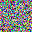}}       
& \raisebox{-.5\height}{\includegraphics[scale=1.3]{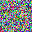}}       \\
20    
& \raisebox{-.5\height}{\includegraphics[scale=1.3]{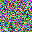}}       
& \raisebox{-.5\height}{\includegraphics[scale=1.3]{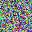}}       
& \raisebox{-.5\height}{\includegraphics[scale=1.3]{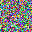}}       
& \raisebox{-.5\height}{\includegraphics[scale=1.3]{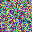}}       \\
40    
& \raisebox{-.5\height}{\includegraphics[scale=1.3]{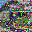}}       
& \raisebox{-.5\height}{\includegraphics[scale=1.3]{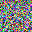}}       
& \raisebox{-.5\height}{\includegraphics[scale=1.3]{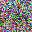}}       
& \raisebox{-.5\height}{\includegraphics[scale=1.3]{imgs/dp_atk/true_40_b.png}}       \\
80    
& \raisebox{-.5\height}{\includegraphics[scale=1.3]{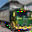}}       
& \raisebox{-.5\height}{\includegraphics[scale=1.3]{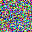}}       
& \raisebox{-.5\height}{\includegraphics[scale=1.3]{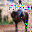}}       
& \raisebox{-.5\height}{\includegraphics[scale=1.3]{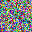}}       \\
120    
& \raisebox{-.5\height}{\includegraphics[scale=1.3]{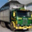}}       
& \raisebox{-.5\height}{\includegraphics[scale=1.3]{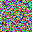}}       
& \raisebox{-.5\height}{\includegraphics[scale=1.3]{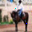}}       
& \raisebox{-.5\height}{\includegraphics[scale=1.3]{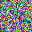}}     \\  \bottomrule
\end{tabular}
\end{center}
\end{table}

\textbf{Potential Trade-off}. The proposed framework provides a series of strategies for implementing encoded gradient aggregation in federated learning. Furthermore, the efficiency of the encoder-decoder network depends on several factors that decide the encoding-decoding error. 1) \textbf{The architecture of the encoder-decoder network}. The size of the encoder-decoder network effect the encoding-decoding error. Specifically, a good architecture with a larger network could result in a lower encoding-decoding error and require more computation. Furthermore, we could use an asymmetrical encoder-decoder (larger decoder, smaller encoder) to decrease the client computation burden. 2) \textbf{Quantization level}. According to our analysis and experiment demonstration, the larger quantization level $s$ could reduce the encoding-decoding error. However, a larger quantization level $s$ leads to a huge quantization domain $\mathbb{Q}_s$, which makes the pre-training stage costs more resources (time and computation). 3) \textbf{Communication efficiency level}. As we demonstrate in Table \ref{tb:ce}, a pre-trained encoder-decoder network with communication efficient benefits will sacrifice the convergence performance to some extent. These potential trade-offs suggest a better-designed architecture of the encoder-decoder network for EGA. Furthermore, the study about the communication efficiency of EGA reveals its potential in gradient compression. We leave these study points in future work.

\textbf{Privacy Risk}. Though EGA is an approximation framework of homomorphic addition, their privacy-preserving ability is not equal. The paper~\citep{DBLP:journals/corr/abs-2001-02610} has proved that noisy gradients, compression, and sparsification could prevent privacy leakage. Hence, EGA could provide specific protection via encoding-decoding error as we evaluated. Besides, we note that if the distribution of the error to Gaussian distribution, then EGA implements differential privacy in an alternative way. We will study the differential privacy mechanism for EGA in the future.



\section{Conclusion}
In this paper, we propose a practical framework EGA, which enables gradients to be aggregated in an encoded domain via pre-trained encode-decode networks. Furthermore, we theoretically prove that the corresponding algorithm has a convergence guarantee and the encoding-decoding errors are bounded. Our empirical results demonstrate that EGA is adaptive to different FL settings while maintaining the final performance of FL models. The gradient protection study also shows that EGA can prevent gradients leakage.

\bibliographystyle{unsrtnat}
\bibliography{main.bib}

\begin{thebibliography}{36}
\providecommand{\natexlab}[1]{#1}
\providecommand{\url}[1]{\texttt{#1}}
\expandafter\ifx\csname urlstyle\endcsname\relax
  \providecommand{\doi}[1]{doi: #1}\else
  \providecommand{\doi}{doi: \begingroup \urlstyle{rm}\Url}\fi

\bibitem[McMahan et~al.(2017)McMahan, Moore, Ramage, Hampson, and
  y~Arcas]{DBLP:conf/aistats/McMahanMRHA17}
Brendan McMahan, Eider Moore, Daniel Ramage, Seth Hampson, and
  Blaise~Ag{\"{u}}era y~Arcas.
\newblock Communication-efficient learning of deep networks from decentralized
  data.
\newblock In Aarti Singh and Xiaojin~(Jerry) Zhu, editors, \emph{Proceedings of
  the 20th International Conference on Artificial Intelligence and Statistics,
  {AISTATS} 2017, 20-22 April 2017, Fort Lauderdale, FL, {USA}}, volume~54 of
  \emph{Proceedings of Machine Learning Research}, pages 1273--1282. {PMLR},
  2017.

\bibitem[Kairouz et~al.(2019)Kairouz, McMahan, Avent, Bellet, Bennis, Bhagoji,
  Bonawitz, Charles, Cormode, Cummings, et~al.]{kairouz2019advances}
Peter Kairouz, H~Brendan McMahan, Brendan Avent, Aur{\'e}lien Bellet, Mehdi
  Bennis, Arjun~Nitin Bhagoji, Kallista Bonawitz, Zachary Charles, Graham
  Cormode, Rachel Cummings, et~al.
\newblock Advances and open problems in federated learning.
\newblock \emph{arXiv preprint arXiv:1912.04977}, 2019.

\bibitem[Yang et~al.(2019)Yang, Liu, Chen, and
  Tong]{DBLP:journals/tist/YangLCT19}
Qiang Yang, Yang Liu, Tianjian Chen, and Yongxin Tong.
\newblock Federated machine learning: Concept and applications.
\newblock \emph{{ACM} Trans. Intell. Syst. Technol.}, 10\penalty0 (2):\penalty0
  12:1--12:19, 2019.

\bibitem[Bhagoji et~al.(2019)Bhagoji, Chakraborty, Mittal, and
  Calo]{DBLP:conf/icml/BhagojiCMC19}
Arjun~Nitin Bhagoji, Supriyo Chakraborty, Prateek Mittal, and Seraphin~B. Calo.
\newblock Analyzing federated learning through an adversarial lens.
\newblock In Kamalika Chaudhuri and Ruslan Salakhutdinov, editors,
  \emph{Proceedings of the 36th International Conference on Machine Learning,
  {ICML} 2019, 9-15 June 2019, Long Beach, California, {USA}}, volume~97 of
  \emph{Proceedings of Machine Learning Research}, pages 634--643. {PMLR},
  2019.

\bibitem[Zhao et~al.(2020)Zhao, Mopuri, and
  Bilen]{DBLP:journals/corr/abs-2001-02610}
Bo~Zhao, Konda~Reddy Mopuri, and Hakan Bilen.
\newblock idlg: Improved deep leakage from gradients.
\newblock \emph{CoRR}, abs/2001.02610, 2020.

\bibitem[Zhu and Han(2020)]{DBLP:series/lncs/Zhu020}
Ligeng Zhu and Song Han.
\newblock Deep leakage from gradients.
\newblock In \emph{Federated Learning}, volume 12500 of \emph{Lecture Notes in
  Computer Science}, pages 17--31. Springer, 2020.

\bibitem[Yin et~al.(2021)Yin, Mallya, Vahdat, Alvarez, Kautz, and
  Molchanov]{DBLP:conf/cvpr/YinMVAKM21}
Hongxu Yin, Arun Mallya, Arash Vahdat, Jose~M. Alvarez, Jan Kautz, and Pavlo
  Molchanov.
\newblock See through gradients: Image batch recovery via gradinversion.
\newblock In \emph{{IEEE} Conference on Computer Vision and Pattern
  Recognition, {CVPR} 2021, virtual, June 19-25, 2021}, pages 16337--16346.
  Computer Vision Foundation / {IEEE}, 2021.

\bibitem[Yao(1986)]{mpcyao}
Andrew Chi-Chih Yao.
\newblock How to generate and exchange secrets.
\newblock In \emph{27th Annual Symposium on Foundations of Computer Science
  (sfcs 1986)}, pages 162--167, 1986.
\newblock \doi{10.1109/SFCS.1986.25}.

\bibitem[Bonawitz et~al.(2017)Bonawitz, Ivanov, Kreuter, Marcedone, McMahan,
  Patel, Ramage, Segal, and Seth]{bonawitz2017practical}
Keith Bonawitz, Vladimir Ivanov, Ben Kreuter, Antonio Marcedone, H~Brendan
  McMahan, Sarvar Patel, Daniel Ramage, Aaron Segal, and Karn Seth.
\newblock Practical secure aggregation for privacy-preserving machine learning.
\newblock In \emph{proceedings of the 2017 ACM SIGSAC Conference on Computer
  and Communications Security}, pages 1175--1191, 2017.

\bibitem[Gentry(2009)]{gentry2009fully}
Craig Gentry.
\newblock Fully homomorphic encryption using ideal lattices.
\newblock In \emph{Proceedings of the forty-first annual ACM symposium on
  Theory of computing}, pages 169--178, 2009.

\bibitem[Paillier(1999)]{paillier1999public}
Pascal Paillier.
\newblock Public-key cryptosystems based on composite degree residuosity
  classes.
\newblock In \emph{International conference on the theory and applications of
  cryptographic techniques}, pages 223--238. Springer, 1999.

\bibitem[Zhang et~al.(2020)Zhang, Li, Xia, Wang, Yan, and
  Liu]{DBLP:conf/usenix/ZhangLX00020}
Chengliang Zhang, Suyi Li, Junzhe Xia, Wei Wang, Feng Yan, and Yang Liu.
\newblock Batchcrypt: Efficient homomorphic encryption for cross-silo federated
  learning.
\newblock In Ada Gavrilovska and Erez Zadok, editors, \emph{2020 {USENIX}
  Annual Technical Conference, {USENIX} {ATC} 2020, July 15-17, 2020}, pages
  493--506. {USENIX} Association, 2020.

\bibitem[Hardy et~al.(2017)Hardy, Henecka, Ivey-Law, Nock, Patrini, Smith, and
  Thorne]{hardy2017private}
Stephen Hardy, Wilko Henecka, Hamish Ivey-Law, Richard Nock, Giorgio Patrini,
  Guillaume Smith, and Brian Thorne.
\newblock Private federated learning on vertically partitioned data via entity
  resolution and additively homomorphic encryption.
\newblock \emph{arXiv preprint arXiv:1711.10677}, 2017.

\bibitem[Phong et~al.(2018)Phong, Aono, Hayashi, Wang, and
  Moriai]{DBLP:journals/tifs/PhongAHWM18}
Le~Trieu Phong, Yoshinori Aono, Takuya Hayashi, Lihua Wang, and Shiho Moriai.
\newblock Privacy-preserving deep learning via additively homomorphic
  encryption.
\newblock \emph{{IEEE} Trans. Inf. Forensics Secur.}, 13\penalty0 (5):\penalty0
  1333--1345, 2018.
\newblock \doi{10.1109/TIFS.2017.2787987}.
\newblock URL \url{https://doi.org/10.1109/TIFS.2017.2787987}.

\bibitem[Wei et~al.(2020)Wei, Li, Ding, Ma, Yang, Farokhi, Jin, Quek, and
  Poor]{wei2020federated}
Kang Wei, Jun Li, Ming Ding, Chuan Ma, Howard~H Yang, Farhad Farokhi, Shi Jin,
  Tony~QS Quek, and H~Vincent Poor.
\newblock Federated learning with differential privacy: Algorithms and
  performance analysis.
\newblock \emph{IEEE Transactions on Information Forensics and Security},
  15:\penalty0 3454--3469, 2020.

\bibitem[Lee and Kifer(2018)]{lee2018concentrated}
Jaewoo Lee and Daniel Kifer.
\newblock Concentrated differentially private gradient descent with adaptive
  per-iteration privacy budget.
\newblock In \emph{Proceedings of the 24th ACM SIGKDD International Conference
  on Knowledge Discovery \& Data Mining}, pages 1656--1665, 2018.

\bibitem[Phan et~al.(2019)Phan, Vu, Liu, Jin, Dou, Wu, and
  Thai]{DBLP:conf/ijcai/PhanVLJDWT19}
NhatHai Phan, Minh~N. Vu, Yang Liu, Ruoming Jin, Dejing Dou, Xintao Wu, and
  My~T. Thai.
\newblock Heterogeneous gaussian mechanism: Preserving differential privacy in
  deep learning with provable robustness.
\newblock In Sarit Kraus, editor, \emph{Proceedings of the Twenty-Eighth
  International Joint Conference on Artificial Intelligence, {IJCAI} 2019,
  Macao, China, August 10-16, 2019}, pages 4753--4759. ijcai.org, 2019.
\newblock \doi{10.24963/ijcai.2019/660}.
\newblock URL \url{https://doi.org/10.24963/ijcai.2019/660}.

\bibitem[Li et~al.(2020{\natexlab{a}})Li, Sahu, Zaheer, Sanjabi, Talwalkar, and
  Smith]{DBLP:conf/mlsys/LiSZSTS20}
Tian Li, Anit~Kumar Sahu, Manzil Zaheer, Maziar Sanjabi, Ameet Talwalkar, and
  Virginia Smith.
\newblock Federated optimization in heterogeneous networks.
\newblock In \emph{MLSys}. mlsys.org, 2020{\natexlab{a}}.

\bibitem[Li et~al.(2020{\natexlab{b}})Li, Sanjabi, Beirami, and
  Smith]{DBLP:conf/iclr/LiSBS20}
Tian Li, Maziar Sanjabi, Ahmad Beirami, and Virginia Smith.
\newblock Fair resource allocation in federated learning.
\newblock In \emph{{ICLR}}. OpenReview.net, 2020{\natexlab{b}}.

\bibitem[Karimireddy et~al.(2020)Karimireddy, Kale, Mohri, Reddi, Stich, and
  Suresh]{karimireddy2020scaffold}
Sai~Praneeth Karimireddy, Satyen Kale, Mehryar Mohri, Sashank Reddi, Sebastian
  Stich, and Ananda~Theertha Suresh.
\newblock Scaffold: Stochastic controlled averaging for federated learning.
\newblock In \emph{International Conference on Machine Learning}, pages
  5132--5143. PMLR, 2020.

\bibitem[Acar et~al.(2021)Acar, Zhao, Navarro, Mattina, Whatmough, and
  Saligrama]{DBLP:conf/iclr/AcarZNMWS21}
Durmus Alp~Emre Acar, Yue Zhao, Ramon~Matas Navarro, Matthew Mattina, Paul~N.
  Whatmough, and Venkatesh Saligrama.
\newblock Federated learning based on dynamic regularization.
\newblock In \emph{9th International Conference on Learning Representations,
  {ICLR} 2021, Virtual Event, Austria, May 3-7, 2021}. OpenReview.net, 2021.
\newblock URL \url{https://openreview.net/forum?id=B7v4QMR6Z9w}.

\bibitem[Oh et~al.(2022)Oh, Kim, and Yun]{oh2022fedbabu}
Jaehoon Oh, SangMook Kim, and Se-Young Yun.
\newblock Fed{BABU}: Toward enhanced representation for federated image
  classification.
\newblock In \emph{International Conference on Learning Representations}, 2022.
\newblock URL \url{https://openreview.net/forum?id=HuaYQfggn5u}.

\bibitem[Abrahamyan et~al.(2021)Abrahamyan, Chen, Bekoulis, and
  Deligiannis]{DBLP:journals/corr/abs-2103-08870}
Lusine Abrahamyan, Yiming Chen, Giannis Bekoulis, and Nikos Deligiannis.
\newblock Learned gradient compression for distributed deep learning.
\newblock \emph{CoRR}, abs/2103.08870, 2021.

\bibitem[Li and Han(2019)]{DBLP:conf/dcc/LiH19}
Hongyu Li and Tianqi Han.
\newblock An end-to-end encrypted neural network for gradient updates
  transmission in federated learning.
\newblock In \emph{{DCC}}, page 589. {IEEE}, 2019.

\bibitem[Roy~Chowdhury et~al.(2020)Roy~Chowdhury, Wang, He, Machanavajjhala,
  and Jha]{roy2020crypt}
Amrita Roy~Chowdhury, Chenghong Wang, Xi~He, Ashwin Machanavajjhala, and Somesh
  Jha.
\newblock Crypt$\varepsilon$: Crypto-assisted differential privacy on untrusted
  servers.
\newblock In \emph{Proceedings of the 2020 ACM SIGMOD International Conference
  on Management of Data}, pages 603--619, 2020.

\bibitem[Alistarh et~al.(2017)Alistarh, Grubic, Li, Tomioka, and
  Vojnovic]{alistarh2017qsgd}
Dan Alistarh, Demjan Grubic, Jerry Li, Ryota Tomioka, and Milan Vojnovic.
\newblock Qsgd: Communication-efficient sgd via gradient quantization and
  encoding.
\newblock \emph{Advances in Neural Information Processing Systems},
  30:\penalty0 1709--1720, 2017.

\bibitem[Zeng et~al.(2021)Zeng, Liang, Hu, and Xu]{zeng2021fedlab}
Dun Zeng, Siqi Liang, Xiangjing Hu, and Zenglin Xu.
\newblock Fedlab: A flexible federated learning framework.
\newblock \emph{arXiv preprint arXiv:2107.11621}, 2021.

\bibitem[Deng(2012)]{deng2012mnist}
Li~Deng.
\newblock The mnist database of handwritten digit images for machine learning
  research [best of the web].
\newblock \emph{IEEE Signal Processing Magazine}, 29\penalty0 (6):\penalty0
  141--142, 2012.

\bibitem[Shakespeare(2014)]{shakespeare2014complete}
William Shakespeare.
\newblock \emph{The complete works of William Shakespeare}.
\newblock Race Point Publishing, 2014.

\bibitem[Caldas et~al.(2018)Caldas, Wu, Li, Kone{\v{c}}n{\'y}, McMahan, Smith,
  and Talwalkar]{DBLP:journals/corr/abs-1812-01097}
Sebastian Caldas, Peter Wu, Tian Li, Jakub Kone{\v{c}}n{\'y}, H.~Brendan
  McMahan, Virginia Smith, and Ameet Talwalkar.
\newblock {LEAF:} {A} benchmark for federated settings.
\newblock \emph{CoRR}, abs/1812.01097, 2018.

\bibitem[Neelakantan et~al.(2015)Neelakantan, Vilnis, Le, Sutskever, Kaiser,
  Kurach, and Martens]{DBLP:journals/corr/NeelakantanVLSK15}
Arvind Neelakantan, Luke Vilnis, Quoc~V. Le, Ilya Sutskever, Lukasz Kaiser,
  Karol Kurach, and James Martens.
\newblock Adding gradient noise improves learning for very deep networks.
\newblock \emph{CoRR}, abs/1511.06807, 2015.
\newblock URL \url{http://arxiv.org/abs/1511.06807}.

\bibitem[Simsekli et~al.(2019)Simsekli, Sagun, and
  Gurbuzbalaban]{simsekli2019tail}
Umut Simsekli, Levent Sagun, and Mert Gurbuzbalaban.
\newblock A tail-index analysis of stochastic gradient noise in deep neural
  networks.
\newblock In \emph{International Conference on Machine Learning}, pages
  5827--5837. PMLR, 2019.

\bibitem[Wu et~al.(2020)Wu, Hu, Xiong, Huan, Braverman, and Zhu]{wu2020noisy}
Jingfeng Wu, Wenqing Hu, Haoyi Xiong, Jun Huan, Vladimir Braverman, and
  Zhanxing Zhu.
\newblock On the noisy gradient descent that generalizes as sgd.
\newblock In \emph{International Conference on Machine Learning}, pages
  10367--10376. PMLR, 2020.

\bibitem[LeCun et~al.(2015)]{lecun2015lenet}
Yann LeCun et~al.
\newblock Lenet-5, convolutional neural networks.
\newblock \emph{URL: http://yann. lecun. com/exdb/lenet}, 20\penalty0
  (5):\penalty0 14, 2015.

\bibitem[Krizhevsky et~al.(2009)Krizhevsky, Hinton,
  et~al.]{krizhevsky2009learning}
Alex Krizhevsky, Geoffrey Hinton, et~al.
\newblock Learning multiple layers of features from tiny images.
\newblock 2009.

\bibitem[Amiri et~al.(2020)Amiri, Gunduz, Kulkarni, and
  Poor]{amiri2020federated}
Mohammad~Mohammadi Amiri, Deniz Gunduz, Sanjeev~R Kulkarni, and H~Vincent Poor.
\newblock Federated learning with quantized global model updates.
\newblock \emph{arXiv preprint arXiv:2006.10672}, 2020.

\end{thebibliography}

\clearpage
\appendix

\section{Proof of Theorem 1} \label{Proofs}

\begin{lemma}\label{lma:boundqda}
The variance of the $k$-th quantified gradients aggregation $\big\| \frac{n}{s} \frac{1}{m}\sum_{i=1}^m Q(\boldsymbol{x}_i^k, s, n) - \frac{1}{m}\sum_{i=1}^m \boldsymbol{x}_i^k \big\|_2^2$ is bounded by $\frac{n}{m}\min(\frac{d}{s^2}, \frac{\sqrt{d}}{s})$.
\end{lemma}
\begin{proof}
We assume $\boldsymbol{x} \in \mathbb{R}^d, 0 \leq k < d$. For all $k$-th element of vectors $(\boldsymbol{x}_1, \boldsymbol{x}_2, ..., \boldsymbol{x}_m)$. The variance bound of quantified vector is given by:
$$
\begin{aligned}
&\mathbb{E}\big(\big\|\ \frac{n}{s} \frac{1}{m}\sum_{i=1}^m Q(\boldsymbol{x}_i^k, s, n) - \frac{1}{m}\sum_{i=1}^m \boldsymbol{x}_i^k \big\|_2^2\big)  \\
&=\frac{1}{m^2} \mathbb{E}\big(\big\|\sum_{i=1}^m(\frac{n}{s} Q(\boldsymbol{x}_i^k, s, n)-\boldsymbol{x}_i^k)\big\|_2^2\big)  \\
&\leq \frac{1}{m^2}\mathbb{E}\big(\sum_{i=1}^m\big\|\frac{n}{s}\boldsymbol{l}_i^k-\boldsymbol{x}_i^k\big\|_2^2\big) \quad \text{(by lemma 3.1 in \cite{alistarh2017qsgd}})\\
&\leq \frac{n}{m^2}\sum_{i=1}^m \min \big(\frac{d}{s^2}, \frac{\sqrt{d}}{s}\big)  \\
&= \frac{n}{m} \min\big(\frac{d}{s^2}, \frac{\sqrt{d}}{s}\big).
\end{aligned}
$$
\end{proof}

\textbf{Proof of Theorem 1}.
\begin{proof} \label{prf:bound:cda}
We assume $\boldsymbol{x} \in \mathbb{R}^d, 0 \leq k < d$. For all $k$-th element of vectors $(\boldsymbol{x}_1, \boldsymbol{x}_2, ..., \boldsymbol{x}_m)$. If noise $\boldsymbol{\nu} \sim \mathcal{N}(0, \sigma^2\mathbf{I})$, then we have the variance bound of EGA:
$$
\begin{aligned}
&\mathbb{E}\Big(\big\|\frac{n}{s} D_c\big(\frac{1}{m} \sum^m_{i=1} E_c\big(Q(\boldsymbol{x}_i^k, s, n)\big)\big) - \frac{1}{m}\sum_{i=1}^m \boldsymbol{x}_i^k \big\|_2^2\Big) \\
&= \mathbb{E}\Big(\big\|\frac{n}{s}(\frac{1}{m}\sum_{i=1}^m \boldsymbol{l}_i^k + \boldsymbol{\nu}^k) - \frac{1}{m}\sum_{i=1}^m \boldsymbol{x}_i^k \big \|_2^2\Big) \qquad\\
&= \mathbb{E}\big(\big\|\frac{1}{m}\sum_{i=1}^m \frac{n}{s} \boldsymbol{l}_i^k - \frac{1}{m}\sum_{i=1}^m \boldsymbol{x}_i^k + \frac{n}{s} \boldsymbol{\nu}^k \big\|_2^2\big) \\
&\leq \mathbb{E}\big(\big\|\frac{1}{m}\sum_{i=1}^m \frac{n}{s} \boldsymbol{l}_i^k - \frac{1}{m}\sum_{i=1}^m \boldsymbol{x}_i^k\big\|_2^2\big) + \mathbb{E}\big(\big\|\frac{n}{s} \boldsymbol{\nu}^k\big\|_2^2\big) \\
&\leq \mathbb{E}\big(\big\|\frac{1}{m}\sum_{i=1}^m \frac{n}{s} \boldsymbol{l}_i^k - \frac{1}{m}\sum_{i=1}^m \boldsymbol{x}_i^k\big\|_2^2\big) + \frac{n^2}{s^2} \mathbb{E}\big( \big\| \boldsymbol{\nu}^k \big\|_2^2\big) \\
&\leq \frac{1}{m^2}\mathbb{E}\big(\sum_{i=1}^m\big\|\frac{n}{s} \boldsymbol{l}_i^k -  \boldsymbol{x}_i^k\big\|_2^2\big) + \frac{n^2}{s^2} \mathbb{E}\big( \big\| \boldsymbol{\nu}^k \big\|_2^2\big)  \qquad (\textrm{by lemma \ref{lma:boundqda}})\\
&\leq \frac{n}{m} \min\big(\frac{d}{s^2}, \frac{\sqrt{d}}{s}\big) + \frac{n^2}{s^2} \sigma^2 .
\end{aligned}
$$
\end{proof}

\section{Convergence Analysis of EGA} \label{convergence}

\subsection{Preliminaries}

We denote the optimal solution minimizing loss function $f(\boldsymbol{w})$ by $\boldsymbol{w}^{*}$, and the minimum loss as $f^{*}$, i.e., $w^{*} \triangleq \arg \min _{w} f(\boldsymbol{w})$, and $f^{*} \triangleq f\big(\boldsymbol{w}^{*}\big)$. And the ServerProcedure updates the global model as
\begin{align}
    \boldsymbol{w}^{t+1}
    &={\boldsymbol{w}}^{t}-\eta_{t} \frac{n}{s} \frac{1}{m} D_c\Big( \sum_{i=1}^{m} E_c \Big( \sum_{j=1}^{k} Q\big(\nabla f_{i}\big(\boldsymbol{w}_{i, j}^{t}, \xi_{i, j}^{t}, s, n\big)\big) \Big) \Big) \\
    &={\boldsymbol{w}}^{t}-\eta_{t} \frac{1}{m} \sum_{i=1}^{m} \sum_{j=1}^{k}  \nabla f_{i}\big(\boldsymbol{w}_{i, j}^{t}, \xi_{i, j}^{t}\big)- \eta_{t}\frac{n}{s}\nu \\
    &={\boldsymbol{w}}^{t}-\frac{1}{m}  \sum_{i=1}^{m} \Delta \boldsymbol{w}_{i}^{t} -\eta_{t} \frac{n}{s}\nu.
\end{align}

\begin{assumption} [$\mu$-Strongly convex, $L$-Smoothness, Bounded variance] \label{assumption:1}
\;
\begin{itemize}
    \item[(a)] The loss functions $f_{1}, \ldots, f_{i}$ are $\mu$-strongly convex; that is, $\forall \boldsymbol{v}, \boldsymbol{u} \in \mathbb{R}^{d}$, i.e, $2\big(f_{i}(\boldsymbol{v})-f_{u}(\boldsymbol{u})\big) \geq 2\big\langle\boldsymbol{v}-\boldsymbol{u}, \nabla f_{i}(\boldsymbol{u})\big\rangle+\mu\|\boldsymbol{v}-\boldsymbol{u}\|_{2}^{2}, \quad \forall i \in[m]$ .
    
    \item[(b)] The loss functions $f_{1}, \ldots, f_{i}$ are $L$-smooth; that is, $\forall \boldsymbol{v}, \boldsymbol{u} \in \mathbb{R}^{d}$, i.e, $2\big(f_{i}(\boldsymbol{v})-f_{i}(\boldsymbol{u})\big) \leq 2\big\langle\boldsymbol{v}-\boldsymbol{u}, \nabla f_{i}(\boldsymbol{u})\big\rangle+L\|\boldsymbol{v}-\boldsymbol{u}\|_{2}^{2}, \quad \forall i \in[m]$ .
    
    \item[(c)] The expected squared $l_{2}$-norm of the stochastic gradients are bounded, i.e., $\mathbb{E}_{\xi}\Big[\big\|\nabla f_{i}\big(\boldsymbol{w}_{i, j}^{t}, \xi_{i, j}^{t}\big)\big\|_{2}^{2}\Big] \leq G^{2}, \quad \forall i \in[k], \forall i \in[m], \forall t$ .

\end{itemize}
\end{assumption}



\subsection{Convergence Rate}
\begin{theorem} \label{them:Convergence Rate}
Let $0<\eta_{t} \leq \min \big\{1, \frac{1}{\mu k}\big\}, \forall t$. We have
\begin{align} \label{eq:Convergence Rate}
    \mathbb{E}\Big[\big\|\boldsymbol{w}^{t}-\boldsymbol{w}^{*}\big\|_{2}^{2}\Big] \leq\Big(\prod_{i=0}^{t-1} A(i)\Big)\big\|\boldsymbol{w}(0)-\boldsymbol{w}^{*}\big\|_{2}^{2}+\sum_{i=0}^{t-1} B(j) \prod_{i=j+1}^{t-1} A(i),
\end{align}
where
\begin{align}
A(i) \triangleq & 1-\mu \eta_{i}\big(k-\eta_{i}(k-1)\big)  \label{eq:A(i)}\\
B(i) \triangleq & {\eta_{t}}^{2}\big(k^{2}+k-1\big) G^{2} + \eta_{t}^2\frac{n^2}{s^2}d \sigma^2\nonumber +\big(1+\mu(1-\eta_{i})\big) \eta^{2}_{i} G^{2} \frac{k(k-1)(2 k-1)}{6}+2 \eta_{i}(k-1) \Gamma. \label{eq:B(i)}
\end{align}
\end{theorem}

\begin{corollary}
From the $L$-smoothness of the loss function, for $0<\eta_{t} \leq \min \left\{1, \frac{1}{\mu k}\right\}, \forall t$, and a total of $T$ global iterations, it follows that

\begin{align}
\mathbb{E}[f(\boldsymbol{w}^{T})]-f^{*} & \leq \frac{L}{2} \mathbb{E}\left[\left\|\boldsymbol{w}^{T}-\boldsymbol{w}^{*}\right\|_{2}^{2}\right] \nonumber\\
& \leq \frac{L}{2}\left(\prod_{i=0}^{T-1} A(i)\right)\left\|\boldsymbol{w}^{0}-\boldsymbol{w}^{*}\right\|_{2}^{2}+\frac{L}{2} \sum_{j=0}^{T-1} B(j) \prod_{i=j+1}^{T-1} A(i),
\end{align}

where the last inequality follows from (\ref{eq:Convergence Rate}). Considering $\eta_{t}=\eta$ and $k=1$, we have
\begin{align}
\mathbb{E}[f(\boldsymbol{w}^{T})]-f^{*} \leq & \frac{L}{2}(1-\mu \eta)^{T}\left\|\boldsymbol{w}^{0}-\boldsymbol{w}^{*}\right\|_{2}^{2} \nonumber +\frac{\eta L}{2\mu}\left(G^{2}+ \frac{n^2}{s^2}d\right)\left(1-(1-\mu \eta)^{T}\right) .
\end{align}
\end{corollary}

\paragraph{Asymptotic convergence} Here we show that, for a decreasing learning rate over time, such that $\lim _{t \rightarrow \infty} \eta_{t}=0$,  $\lim _{t \rightarrow \infty} \mathbb{E}[f(\boldsymbol{w}^{t})]-f^{*}=$ 0 . For $0<\eta_{t} \leq \min \left\{1, \frac{1}{\mu k}\right\}$, we have $0 \leq A(t)<1$, and $\lim _{t \rightarrow \infty} \prod_{i=0}^{t-1} A(i)=0$. For simplicity, assume $\eta_{t}=\frac{\alpha}{t+\beta}$, for constant values $\alpha$ and $\beta .$ For $j \gg 0, B(j) \rightarrow 0$, and for limited $j$ values, $\prod_{i=j+1}^{t-1} A(i) \rightarrow 0$, and so, according to $(12), \lim _{t \rightarrow \infty} \mathbb{E}[f(\boldsymbol{w}^{t})]-f^{*}=0 .$

\subsection{Proof of Theorem \ref{them:Convergence Rate}}
\begin{proof}
We have
\begin{align} \label{eq:toal}
    \mathbb{E}\Big[\big\|\boldsymbol{w}^{t+1}-\boldsymbol{w}^{*}\big\|_{2}^{2}\Big] \nonumber 
    &= \mathbb{E}\Big[\big\|{\boldsymbol{w}}^{t}-\boldsymbol{w}^{*}\big\|_{2}^{2}\Big]+\mathbb{E}\Big[\big\| \frac{1}{m} \sum_{i=1}^{m} \Delta \boldsymbol{w}_{i}^{t} + \eta_{t} \frac{n}{s} \nu \big\|_{2}^{2}\Big] +2 \mathbb{E}\Big[\big\langle{\boldsymbol{w}}^{t}-\boldsymbol{w}^{*}, \frac{1}{m}\sum_{i=1}^{m}  \Delta \boldsymbol{w}_{i}^{t}+\eta_{t} \nu \big\rangle\Big] \nonumber\\
    &\leq \mathbb{E}\Big[\big\|{\boldsymbol{w}}^{t}-\boldsymbol{w}^{*}\big\|_{2}^{2}\Big]+\mathbb{E}\Big[\big\|\frac{1}{m} \sum_{i=1}^{m} \Delta \boldsymbol{w}_{i}^{t} \big\|_{2}^{2}\Big]+
    \mathbb{E}\Big[\big\|{\eta_{t} \frac{n}{s} \nu}\big\|_{2}^{2}\Big] +2 \mathbb{E}\Big[\big\langle{\boldsymbol{w}}^{t}-\boldsymbol{w}^{*}, \frac{1}{m}\sum_{i=1}^{m}  \Delta \boldsymbol{w}_{i}^{t} \big\rangle\Big] .
\end{align}

From the convexity of $\|\cdot\|_{2}^{2}$, it follows that

\begin{align} \label{eq:2nd_term}
\mathbb{E}\bigg[\Big\| \frac{1}{m}\sum_{i=1}^{m}  \Delta \boldsymbol{w}_{i}^{t} \Big\|_{2}^{2}\bigg] 
& \leq \frac{1}{m}\sum_{i=1}^{m}  \mathbb{E}\Big[\big\|\Delta \boldsymbol{w}_{i}^{t}\big\|_{2}^{2}\Big] \nonumber\\
&={\eta_{t}}^{2} \frac{1}{m}\sum_{i=1}^{m}  \mathbb{E}\Big[\big\|\sum_{j=1}^{k} \nabla f_{i}\big(\boldsymbol{w}_{i, j}^{t}, \xi_{i, j}^{t}\big)\big\|_{2}^{2}\Big] \nonumber\\
& \leq {\eta_{t}}^{2} k \frac{1}{m}\sum_{i=1}^{m} \sum_{j=1}^{k}  \mathbb{E}\Big[\big\|\nabla f_{i}\big(\boldsymbol{w}_{i, j}^{t}, \xi_{i, j}^{t}\big)\big\|_{2}^{2}\Big] \leq {\eta_{t}}^{2} k^{2} G^{2}, \qquad \text{(by Assumption \ref{assumption:1})}
\end{align}

and
\begin{align} \label{eq:3rd_term}
\mathbb{E}\Big[\big\|{\eta_{t} \frac{s}{n} \nu}\big\|_{2}^{2}\Big] \leq \eta_{t}^2 \frac{n^2}{s^2} \mathbb{E}\Big[ \sum_{j=1}^{d}\nu_j^2 \Big] \leq 
\eta_{t}^2 \frac{n^2}{s^2} d\sigma^2.
\end{align}

We rewrite the third term on the RHS of (\ref{eq:toal}) as follows:

\begin{align} \label{eq:4th_term}
2 \mathbb{E}\Big[\big\langle\boldsymbol{w}^{t}-\boldsymbol{w}^{*},  \frac{1}{m} \sum_{i=1}^{m} \Delta \boldsymbol{w}_{i}^{t}\big\rangle\Big] 
=& 2 \eta_{t} \frac{1}{m} \sum_{i=1}^{m} \mathbb{E}\bigg[\big\langle\boldsymbol{w}^{*}-\boldsymbol{w}^{t}, \sum_{j=1}^{k} \nabla f_{i}\big(\boldsymbol{w}_{i, j}^{t}, \xi_{i, j}^{t}\big)\big\rangle\bigg] \nonumber\\
=& 2 \eta_{t} \frac{1}{m} \sum_{i=1}^{m} \mathbb{E}\Big[\big\langle\boldsymbol{w}^{*}-\boldsymbol{w}^{t}, \nabla f_{i}\big(\boldsymbol{w}^{t}, \xi_{m, 1}^{t}\big)\big\rangle\Big] \nonumber\\
&+2 \eta_{t} \frac{1}{m} \sum_{i=1}^{m} \mathbb{E}\bigg[\big\langle\boldsymbol{w}^{*}-\boldsymbol{w}^{t}, \sum_{j=2}^{k} \nabla f_{i}\big(\boldsymbol{w}_{i, j}^{t}, \xi_{i, j}^{t}\big)\big\rangle\bigg] .
\end{align}

To bound this term further, we need the following result. We observe the following:
\begin{align} \label{eq:4th_term_eq1}
2 \eta_{t} \frac{1}{m} \sum_{i=1}^{m} \mathbb{E} &\Big[\big\langle\boldsymbol{w}^{*}-\boldsymbol{w}^{t}, \nabla f_{i}\big(\boldsymbol{w}^{t}, \xi_{m, 1}^{t}\big)\big\rangle\Big] \nonumber\\
& = 2 \eta_{t} \frac{1}{m} \sum_{i=1}^{m} \mathbb{E}\Big[\big\langle\boldsymbol{w}^{*}-\boldsymbol{w}^{t}, \nabla f_{i}\big(\boldsymbol{w}^{t}\big)\big\rangle\Big] \qquad (\text{since} \mathbb{E}_{\xi}\big[\nabla f_{i}\big(\boldsymbol{w}_{i, j}^{t}, \xi_{i, j}^{t}\big)\big]=\nabla f_{i}\big(\boldsymbol{w}_{i, j}^{t}\big), \forall i, m )\nonumber\\
& \leq 2 \eta_{t} \frac{1}{m} \sum_{i=1}^{m} \mathbb{E}\Big[f_{i}\big(\boldsymbol{w}^{*}\big)-f_{i}(\boldsymbol{w}^{t})-\frac{\mu}{2}\big\|\boldsymbol{w}^{t}-\boldsymbol{w}^{*}\big\|_{2}^{2}\Big] \qquad (\text{by Assumption \ref{assumption:1}})
\nonumber\\
&=2 \eta_{t}\bigg(f^{*}-\mathbb{E}[f(\boldsymbol{w}^{t})]-\frac{\mu}{2} \mathbb{E}\Big[\big\|\boldsymbol{w}^{t}-\boldsymbol{w}^{*}\big\|_{2}^{2}\Big]\bigg) 
\leq-\mu \eta_{t} \mathbb{E}\Big[\big\|\boldsymbol{w}^{t}-\boldsymbol{w}^{*}\big\|_{2}^{2}\Big] \quad (\text{since} \; f^{*} \leq f\big(\boldsymbol{w}^{t}\big), \forall t)
\end{align}

\begin{lemma}[\cite{amiri2020federated}, Lemma 2] \label{lem: lemma_term2}
For $0\leq \eta_{t} \leq1$
\begin{align} \label{eq:lemma_termE}
2 \eta_{t} \frac{1}{m} \sum_{i=1}^{m} \mathbb{E}\Big[\big\langle\boldsymbol{w}^{*}-\boldsymbol{w}^{t}, \sum_{j=2}^{k} \nabla f_{i}\big(\boldsymbol{w}_{i, j}^{t}, \xi_{i, j}^{t}\big)\big\rangle\Big] 
\leq &-\mu \eta_{t}(1-\eta_{t})(k-1) \mathbb{E}\big[\big\|\boldsymbol{w}^{t}-\boldsymbol{w}^{*}\big\|_{2}^{2}\big]+{\eta_{t}}^{2}(k-1) G^{2} \nonumber\\
&+\big(1+\mu(1-\eta_{t})\big) {\eta_{t}}^{2} G^{2} \frac{k(k-1)(2 k-1)}{6}+2 \eta_{t}(k-1) \Gamma .
\end{align}
\end{lemma}

Substituting the bound obtained from above lemma  (\ref{eq:4th_term_eq1}) and (\ref{eq:lemma_termE}) in Equation (\ref{eq:4th_term}), we obtain:
\begin{align} \label{eq:4th_term_eq2}
2 \mathbb{E}\Big[\big\langle{\boldsymbol{w}}^{t}-\boldsymbol{w}^{*}, \frac{1}{m} \sum_{i=1}^{m} \Delta \boldsymbol{w}_{i}^{t}\big\rangle\Big] \nonumber
& \leq-\mu \eta_{t}(k-\eta_{t}(k-1)) \mathbb{E}\Big[\big\|\boldsymbol{w}^{t}-\boldsymbol{w}^{*}\big\|_{2}^{2}\Big]+{\eta_{t}}^{2}(k-1) G^{2} \nonumber\\
&\quad+(1+\mu(1-\eta_{t})) {\eta_{t}}^{2} G^{2} \frac{k(k-1)(2 k-1)}{6}+2 \eta_{t}(k-1) \Gamma.
\end{align}

Substituting the above inequality in Equation (\ref{eq:2nd_term}), leads to the following upper bound on $\mathbb{E}\big[\big\|\boldsymbol{w}^{t+1}-\boldsymbol{w}^{*}\big\|_{2}^{2}\big]$, when substituted into (\ref{eq:toal}):

\begin{align} 
\mathbb{E}\Big[\big\|\boldsymbol{w}^{t+1}-\boldsymbol{w}^{*}\big\|_{2}^{2}\Big] \leq &\big(1-\mu \eta_{t}(k-\eta_{t}(k-1))\big) \mathbb{E}\Big[\big\|{\boldsymbol{w}}^{t}-\boldsymbol{w}^{*}\big\|_{2}^{2}\Big]+{\eta_{t}}^{2}\big(k^{2}+k-1\big) G^{2} + \eta_{t}^2d \sigma^2 \nonumber\\
&+\big(1+\mu(1-\eta_{t})\big) {\eta_{t}}^{2} G^{2} \frac{k(k-1)(2 k-1)}{6}+2 \eta_{t}(k-1) \Gamma.
\end{align}
\end{proof}

\end{document}